# Investigation of supramolecular architectures of bent-shaped pyridine derivatives: from a three-ring crystalline compound towards five-ring mesogens

Nemanja Trišović,*[a] Jelena Antanasijević,[a] Jelena Rogan,[a] Dejan Poleti,[a] Tibor Tóth-Katona,[b] Miroslaw Salamonczyk,[c] Antal Jákli,[bc] and Katalin Fodor-Csorba[b]

In searching for novel photoactive liquid crystals, we have synthesized a series of five-ring pyridine-based bent-core compounds bearing different substituents at the peripheral phenyl rings ($CH_3O$, Cl and $NO_2$). Their mesomorphic behaviour has been investigated by polarizing optical microscopy, differential scanning calorimetry and X-ray scattering, and then compared with the unsubstituted parent compound. The introduction of the methoxy groups at the peripheral phenyl rings of the bent core results in a non-mesomorphic compound, whereas the chloro- and nitro-substituted compounds form enantiotropic B1-like phases. Significant changes of the textures and transition temperatures of the mesophase have been observed under UV light. The presented investigation of the mesomorphic properties of the synthesized compounds, coupled with analysis of the molecular packing of the related three-ring compounds, will help to design self-organized molecules suitable for UV indicators.

## Introduction

The rational design and structural control of supramolecular systems allow a fine tuning of their macroscopic properties. Pyridine derivatives are especially interesting building blocks for this purpose due to their ability to establish various non-covalent interactions. Additionally, pyridine itself, which is a highly electron deficient aromatic moiety, can be used as the electron-withdrawing part in push–pull systems. This feature opens up opportunities for obtaining photoactive materials with high quantum efficiency as well as good charge transporting ability.

A number of pyridine-based mesogens has been reported in the literature, including rod-like molecules[1–6], bent-shaped molecules with up to five aromatic rings [7–13] and polymers[14–16]. Their mesogenic properties may be attributed to the enhanced unsaturation and polarizability caused by the presence of this heteroaromatic ring[8]. Non-mesogenic pyridine derivatives have been used in mixtures with non-mesogenic carboxylic acids to lead to hydrogen-bonded complexes, which arrange to form different mesophases[17–19]. Halogen bonding has been similarly applied as a directing force for the formation of supramolecular mesogenic complexes between halogens or polysubstituted halobenzenes and pyridine derivatives[20,21]. Furthermore, N-alkylation of the pyridine unit leads to materials, which combine the properties of both liquid crystals and ionic liquids[22–25]. Although pyridine-based mesogens exhibit more complex mesomorphic behaviour than their benzene analogues[5,10], their relatively high transition temperatures limit practical applications. We have demonstrated that, in bent-core mesogens with pyridine at the central ring the mesophase range becomes lower and wider as the polarity of the bent-core is decreased by changing the linking group from amide to ester and to olefinic[9].

In continuation of our work, here we present the synthesis and characterization of a series of five-phenyl ring containing bent-core compounds (**B5**), where we have substituted the hydrogen atom at the peripheral phenyl rings of the bent core **B5-H** bearing the olefinic linking group between the central pyridine ring and wings (Figure 1a)[9], with substituents **X**. In the synthesized compounds the substituent X varies as: **B5-MeO** (X = $CH_3O$), **B5-Cl** (X = Cl) and **B5-NO₂** (X = $NO_2$) (Figure 1b). Due to combination of steric effects and dipolar interactions, peripheral substituents can have a significant effect on the behaviour of bent-shaped mesogens[26]. For example, they can induce ferroelectric behaviour instead of antiferroelectric order[27]. The investigation of the mesomorphic behaviour of the five-ring bent-shaped compounds synthesized here in correlation with the analysis of the crystal structure of the bent-core three-ring precursor (**B3**, Figure 1c) will enable the strategic construction of specific supramolecular systems as well as a more complete understanding of their structure–property relationships.

[a.] Faculty of Technology and Metallurgy, University of Belgrade, Belgrade, Serbia. E-mail: ntrisovic@tmf.bg.ac.rs
[b.] Wigner Research Centre for Physics, Institute for Solid State Physics and Optics of the Hungarian Academy of Sciences, H-1525 Budapest, P.O. Box 49, Hungary
[c.] Liquid Crystal Institute, Kent State University, Kent, Ohio 44242, USA

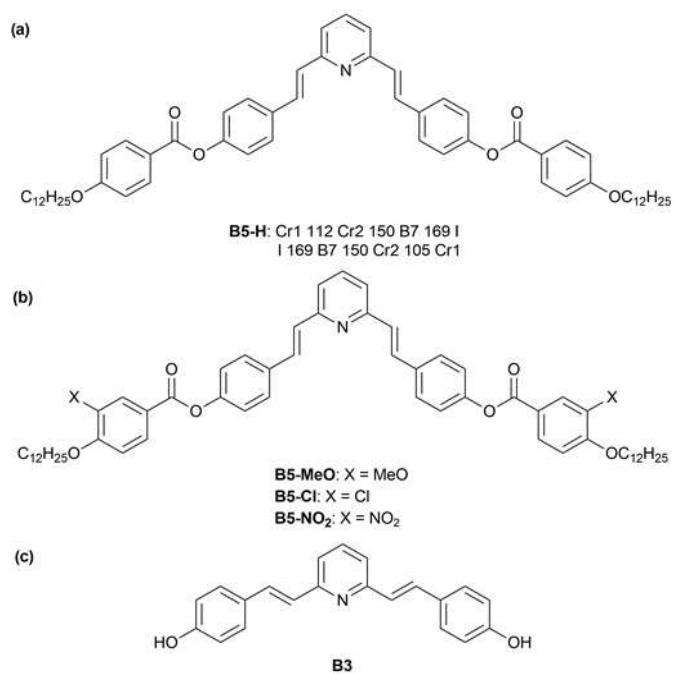

**Fig. 1** (a) Chemical structure of the parent compound **B5-H** and its phase sequence (phase transition temperatures)[11]; (b) and (c) Chemical structures of the investigated compounds

## Results and discussion

### Synthesis

The synthesis of 2,6-bis[2-(4-(4-dodecyloxy-3-substituted benzoyloxy)phenyl)ethenyl]pyridines with three different substituents, methoxy (**B5–MeO**), chloro (**B5–Cl**) and nitro (**B5–NO₂**), in position 3 of the peripheral phenyl rings is presented in Scheme 1. As reported in our previous work[9], 2,6-bis[2-(4-hydroxyphenyl)ethenyl]pyridine (**B3**) was prepared according to the procedure described by Bergmann and Pinchas[28]. Condensation of 2,6-lutidine (**1**) with excess 4-hydroxybenzaldehyde in acetic anhydride at the reflux and subsequent base catalysed hydrolysis of 2,6-bis[2-(4-ethanoyloxyphenyl)ethenyl]pyridine (**2**) led to the compound **B3**. 4-Dodecyloxy-3-substituted benzoic acids (**3a–3c**) were obtained by undertaking the following steps[29]: protection of the carboxylic group by esterification, introduction of the dodecyl tale and the subsequent deprotection of the carboxylic group by alkaline hydrolysis of esters. The synthesis of the target compounds (**B5–MeO**, **B5–Cl** and **B5–NO₂**) was achieved by acylation of compound **B3** with two equivalents of the appropriately substituted benzoyl chlorides in the presence of triethylamine as base.

### Crystal structure of 2,6-bis[2-(4-hydroxyphenyl)ethenyl]pyridine (B3)

When discussing the structure of 2,6-bis(2-arylethenyl)pyridines, one should take into account three conformational isomers which originate from lack of the free rotation of the arylethenyl moieties around the quasi-single bonds with the central pyridine ring (Figure 2). The molecular structure of **B3** is characteristic for bent-shaped molecules with s-*trans*/s-*trans* conformation (Scheme 1, Figure 3). Although at first sight there is a mirror plane passing through the atoms N1 and C20, the molecule is actually twisted around its long axis. In Table 1, selected geometry parameters of **B3** are compared with literature data for some structurally similar compounds. **B3** is unique among them due to s-*trans*/s-*trans* conformation. As expected, the angle between centroids of aromatic rings (the bending angle) of about 134° is very high, the highest among so far structurally characterized compounds (Table 1).

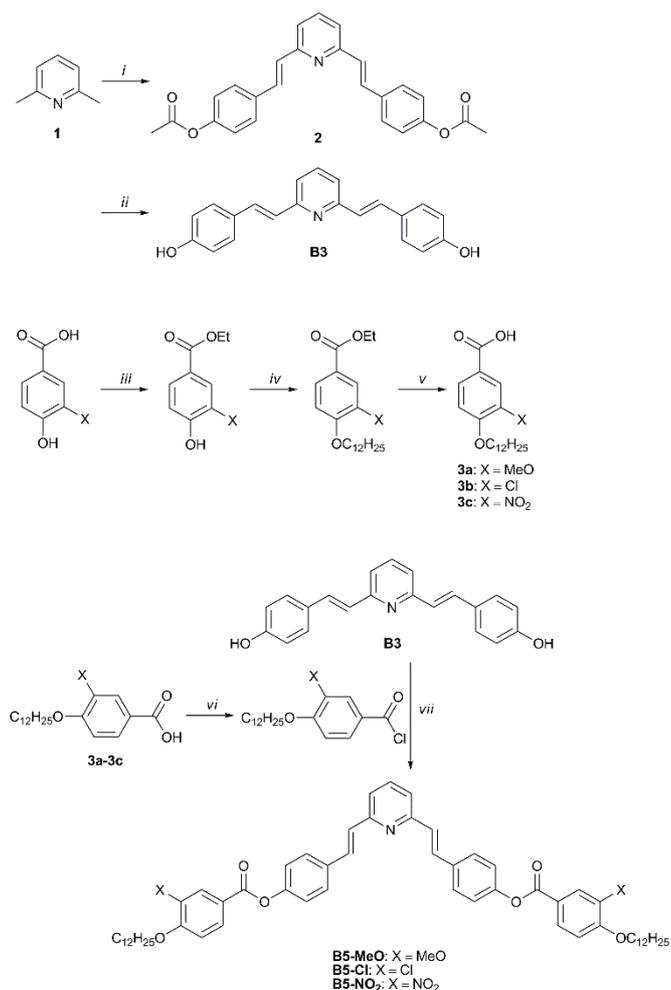

**Scheme 1** Synthesis of the investigated compounds. Reagents and conditions: *i*) 4-OH-C₆H₄-CHO, Ac₂O, reflux, 24 h, (60 %); *ii*) KOH, EtOH, reflux, 1.5 h; then CO₂, RT (32 %); *iii*) EtOH, H₂SO₄, reflux, 24 h (55–60 %); *iv*) C₁₂H₂₅Br, K₂CO₃, DMF, RT, 24 h, (70–80 %); *v*) KOH, 50 % EtOH, reflux, 6 h (70–75 %); *vi*) (COCl)₂, DMF, reflux, 6 h; *vii*) NEt₃, CH₂Cl₂, RT, 48 h, (45–70 %).

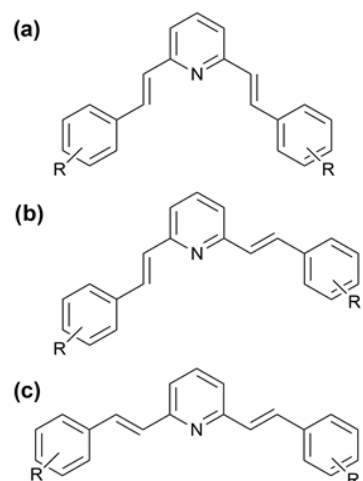

**Fig. 2** Three possible conformations for 2,6-bis[2-arylphenyl]py-ridines: (a) s-*cis*/s-*cis*; (b) s-*cis*/s-*trans*; (c) s-*trans*/s-*trans*.

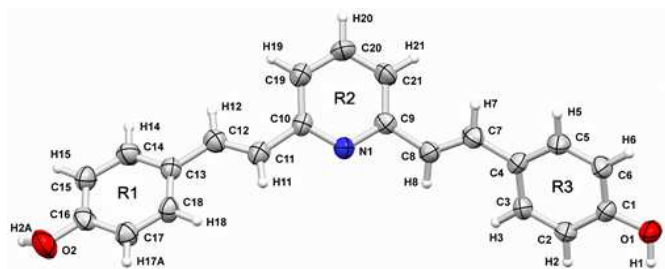

**Fig. 3** The molecular structure of **B3** with atomic numbering scheme and displacement ellipsoids drawn at the 50% probability level. (Isotropic H atoms are shown as spheres of arbitrary radii.)

According to the available crystallographic data, 2,6-bis(2-phenylethenyl)pyridine adopts the s-*cis*/s-*cis* conformation (Table 1), which has been attributed to the presence of a hydrogen-type bond between the central nitrogen and the olefinichydrogens, which makes the bounded conformation**Fig.**

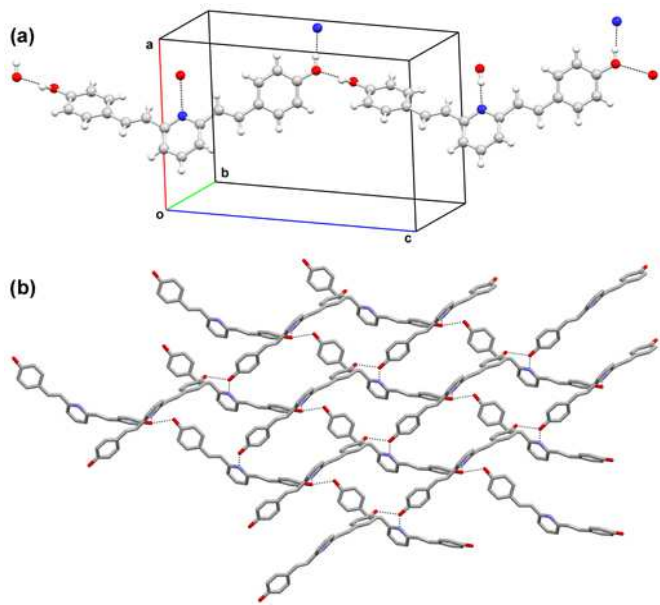

**4** (a) Pseudo-chains of **B3** molecules running approximately along [013] direction. (b) Pseudo-layers of **B3** molecules parallel to the *bc*-plane. Hydrogen bonds are shown as dashed lines.

particularly planar and stable[30,31]. *Ab initio* calculations have confirmed that this conformation is the most stable structure, although the torsion barriers between the three possible conformations are sufficiently small to permit their interconversion in a thermodynamic equilibrium[32].

**Table 1** Selected geometry parameters of **B3** and some structurally related bent-core molecules (see Fig. 2 for labels).

| Compound | Conformation in the solid state | Centroids R1–R2–R3 angle (°) | Dihedral angles between aromatic rings (°) | | | Reference |
|---|---|---|---|---|---|---|
| | | | R1–R2 | R2–R3 | R1–R3 | |
| HO-R1-CH=CH-pyridine-CH=CH-R3-OH | s-*trans*/s-*trans* | 133.7 | 37.5(2) | 34.8(2) | 59.0(2) | this work |
| Ph-CH=CH-pyridine-CH=CH-Ph | s-*cis*/s-*cis* | 95.3 | 10.6 | 5.36 | 15.85 | 30 |
| (MeO)₂Ph-CH=CH-pyridine-CH=CH-Ph(OMe)₂ | s-*cis*/s-*trans* | 116.6 | 16.4 | 13.2 | 20.5 | 33 |
| 3,5-bis(styryl)pyridine | s-*cis*/s-*cis* | 112.8 | 35.4 | 35.4 | 70.1 | 34 |

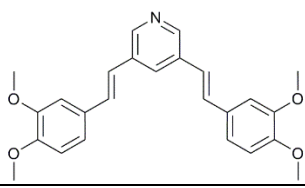

| | | | | | |
|---|---|---|---|---|---|
| | s-*cis*/s-*cis* | 105.7 | 20.3 | 10.9 | 23.4 | 33 |

**Table 2** Geometry of hydrogen bonds

| D–H⋯A | d(D–H) (Å) | d(H⋯A) (Å) | d(D⋯A) (Å) | D–H⋯A (°) |
|---|---|---|---|---|
| O1–H1⋯N1[i] | 0.88 | 1.80 | 2.682(3) | 174 |
| O2–H2A⋯O2[ii] | 0.91 | 1.80 | 2.686(3) | 161 |

Symmetry codes: *i*: –x + 3/2, y + 1/2, –z + 1/2; *ii*: x, y – 1, z – 1.

In the case of **B3**, the hydrogen bonding interactions between the molecules compensate for the energy difference and stabilize the s-*trans*/s-*trans* conformation. When two methoxy groups are introduced at both phenyl groups, only the s-*cis*/s-*trans* isomer is identified in the crystalline sample[33]. It can be summarized here that the deviation from the planarity in 2,6-bis(2-arylethenyl)pyridine increases with increasing the bending angle (Table 1). Interestingly, bent-core molecules with the arylethenyl moieties differently positioned in relation to the pyridine ring nitrogen (3,5-bis(2-arylethenyl)pyridines) have the s-*cis*/s-*cis* conformation in the solid state[33,34]. However, their bending angle is higher and the nonplanarity is larger than the corresponding 2,6-bis(2-arylethenyl)pyridines (Table 1).

For the molecules listed in Table 1, the lengths of the double C=C bonds vary in a very narrow range between about 1.32 and

1.34 Å. Thus, an extended delocalization, which should depend on the conformation of molecules, is not found. Probably the conformation of molecules is governed by crystal packing, although there are no strong, *e.g.* π–π or similar interactions.

Due to the presence of OH groups at both sides of the **B3** molecules, they are connected by O–H⋯O hydrogen bonds (Table 2) making wavy pseudo-chains running approximately along the [013] direction (Figure 4a). At the same time, the pseudo-chains are interconnected by O–H⋯N hydrogen bonds, which are also shown in Figure 4a, making double pseudo-layers parallel to the *bc*-plane and an interesting network structure (Figure 4b). The structurally similar compounds listed in Table 1 do not have any hydrogen-bond donating moiety and their molecules form either stacks or zig-zag chains in the crystal [30,33,34].

**Optical and calorimetric investigations of B5-MeO, B5-Cl and B5-NO₂**

The thermal behaviour of the five-ring compounds **B5-MeO**, **B5-Cl** and **B5-NO₂** has been studied by polarizing optical microscopy (POM) and differential scanning calorimetry (DSC). Their phase sequences, transition temperatures (°C) and associated phase transition enthalpies in kJ mol⁻¹ are summarized in Table 3 and compared to the corresponding values of the parent compound **B5-H**. All compounds are thermally stable as confirmed by the reproducibility of the thermographs in several heating and cooling cycles (Figure 5).

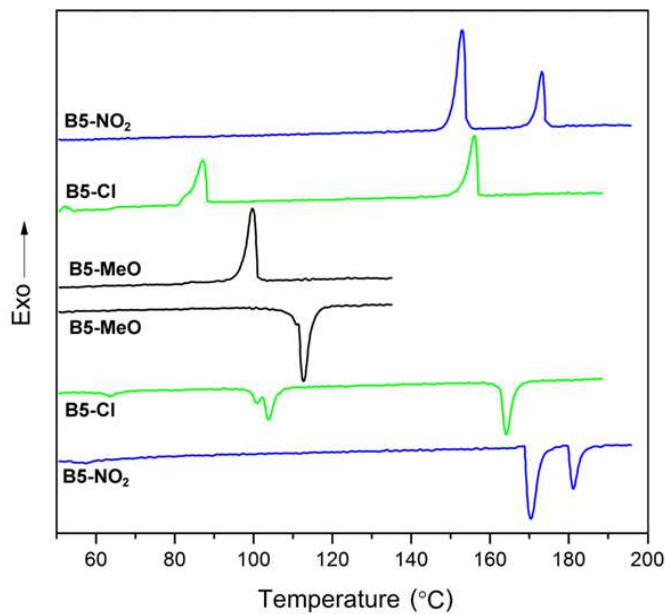

Introduction of the electron-donating methoxy groups at the peripheral phenyl rings of the bent core of **B5-H** results in a non-mesomorphic compound **B5-MeO** having a reduced melting point at 110 °C. On the other hand, the chloro- and nitro-substituted compounds form an enantiotropic B1-like phase characterized by smooth fan-shape domains (Figure 6), which is similar to that of the lower homologue of **B5-H** bearing theeptyloxy tails. In comparison with the parent compound **B5-H**, **Fig. 5** DSC scans for compounds **B5-MeO**, **B5-Cl** and **B5-NO₂** detected at 5 °C min⁻¹ heating/cooling rates. The upper curves correspond to the second cooling and the corresponding lower curves to the second heating run.

themesomorphic range of the chloro-substituted compound **B5-Cl** is significantly broadened, whereas its clearing point remains almost unchanged. The clearing temperature of the compound with the strong electron-withdrawing nitro groups **B5-NO₂** increases by ca. 15 °C, and its mesomorphic range is somewhat narrowed. The observed substituent effect demonstrates that the delicate balance between its size and polarity dictates the packing process.

On the other side, the appearance of a smectic phase indicate that the preferential conformation of these compounds should be *s-cis*/*s-cis*. At high values of bending angles characteristic for the the*s-trans*/s-*trans* conformation (ca. 130°), a nematic phase would appear. In a previous paper[9], we have found that the values of the bending angle of five-ring molecules and three-ring molecules as their structural constituents are in good agreement. Thus, quantum-chemical calculations confirmed that the preferential conformation of the unsubstituted parent compound **B5-H** is also *s-cis*/*s-cis*.

**Small Angle X-ray scattering measurements**

X-ray measurements conformed that **B5-MeO** shows only crystalline phase, which fits with DSC data. SAXS measurements of **B5-Cl** confirmed the presence of an LC phase but in different temperature range than measured by DSC. After the sample was heated to the isotropic phase (clearing point was 163.6 °C in agreement with DSC data) it was cooled at 1 °C min$^{-1}$ cooling rate. The first phase (LC1) was observed in ~3 °C temperature range. The 2-D scattering pattern (see Figure 7a) and the q-dependence of the scattering intensity with the Miller indices (see Figure 7c) of LC1 is consistent with a B1-type modulated smectic phase with a layer spacing of $b$=45.02 Å, modulation periodicity of $a$=121.6 Å and angle between them being $\gamma$ = 112.2°. SAXS pattern of the LC2 phase shows a 3-D periodicity, which however does not seem to be a crystal, as the 2-D scattering pattern is not grainy typical to crystalline powder cells.

**Table 3** Transition temperatures (°C) and the enthalpies of transitions (kJ mol$^{-1}$) for the investigated bent-core mesogens

| Compound | 2$^{nd}$ Heating | 2$^{nd}$ Cooling | Reference |
|---|---|---|---|
| **B5-MeO** | **Cr** 110.9 (–*46.7*) **I** | **I** 102.9 (*36.2*) **Cr** | this work |
| **B5-Cl** | **Cr1** 61.1 (–*2.9*) **Cr2** 102.0 (–*21.6*) **B1** 162.3 (–*18.0*) **I** | **I** 159.3 (*24.1*) **B1** 90.4 (*20.0*) **Cr2** 54.8 **Cr1** | this work |
| **B5-NO$_2$** | **Cr1** 53.5 (–*7.4*) **Cr2** 168.5 (–*36.3*) **B1** 179.3 (–*16.3*) **I** | **I** 176.5 (*18.1*) **B1** 156.3 (*37.8*) **Cr** | this work |
| **B5-H** | **Cr1** 112.2 (–*15.1*) **Cr2** 151.5 (–*41.5*) **B7** 166.8 (–*29.6*) **I** | **I** 159.2 (*32.2*) **B7** 140.2 (*42.5*) **Cr2** 103.5 (–*16.1*) **Cr1** | 9 |

Polarizing optical microscopy studies using 100W XENOPHOT halogen photooptic light bulb (Model: HLX 64623) of Olympus BX60 that emits only above 400 nm, *i.e.*, basically no UV, are shown in the upper row of Figure 8. Textures both at 162 °C (identified by SAXS as a B1-type LC1 phase) and 103.5 °C (denoted as LC2 by SAXS) in (a) and (b), respectively, show columnar (B1) type fan textures. This texture becomes decorated by fine tangential lines below 92 °C remains stable even at room temperature (see Figure 8c at 40 °C) and is clearly not crystalline. The difference between the phase identifications based on DSC, X-ray and POM results indicate different lighting conditions (completely dark during DSC), uncontrolled during SAXS measurements and illuminated by visible light in POM) indicate photoactivity of the material.

To illustrate it further, we have also illuminated a 5 μm **B5-Cl** cell with a 100 W SYLVANIA Mercury spot light (H44GS-100) UV lamp that has a strong I-line at 365.4 nm wavelength during POM studies. For that the lamp was placed at 160 mm distance from the side of the cell. After shining the sample with this UV light for several minutes at 155 °C the transmitted the texture became dark. This dark state recovered even after heating sample to isotropic phase at 165 °C and cooled back to 155 °C.

**Fig. 6** Characteristic textures of **B5-Cl** on cooling at (a) 160 °C and (b) 100 °C; characteristic textures of **B5-NO$_2$** on cooling at (c) 182 °C and (d) 32 °C.

Importantly, the texture remained completely dark even under oblique light, indicating the phase is isotropic, *i.e.*, the clearing point shifted below 155 °C. Cooling the sample to 130 °C abirefringent texture appeared indicating a transition to a mesophase. Such a texture is shown in Figure 8d and it resembles to a B2 phase. Illuminating by the UV light for a second time for about 5 minutes the texture became dark again (see Figure 8e), showing the clearing point dropped below 130 °C. This texture remained basically unchanged (see Figure 8f) even after keeping in complete darkness for 15 hours. Subsequent cooling indicated the occurrence of the mesophase at 120 °C.

Since on both sides of pyridine of **B5-Cl** there are two vinylene units, which can polymerize under UV light, our observations suggest polymerization or decomposition of the material under UV light. The recognized changes in the mesomorphic behaviour of **B5-Cl** might partly be ascribed to this phenomenon. Although this effect is unwanted for certain applications, it can also be used for UV indicators in other applications.

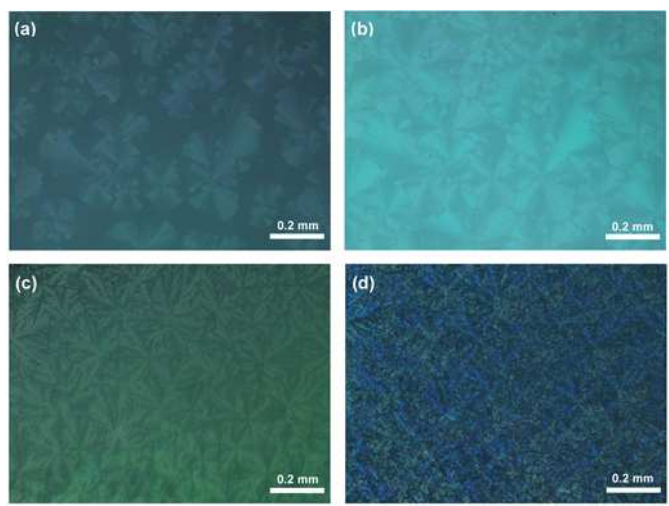

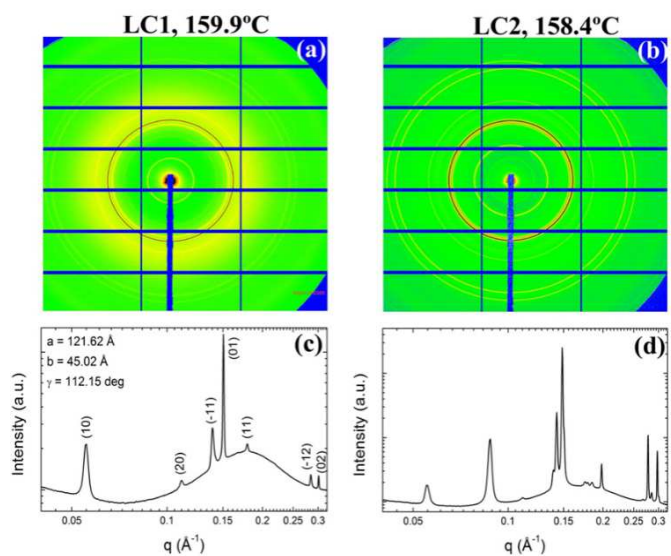

**Fig. 7** Summary of SAXS results obtained on **B5-Cl**. **a** and **b** 2-D scattering profile in the LC1 and LC2 phases; **c** and **d** q-dependences of the scattering intensities.

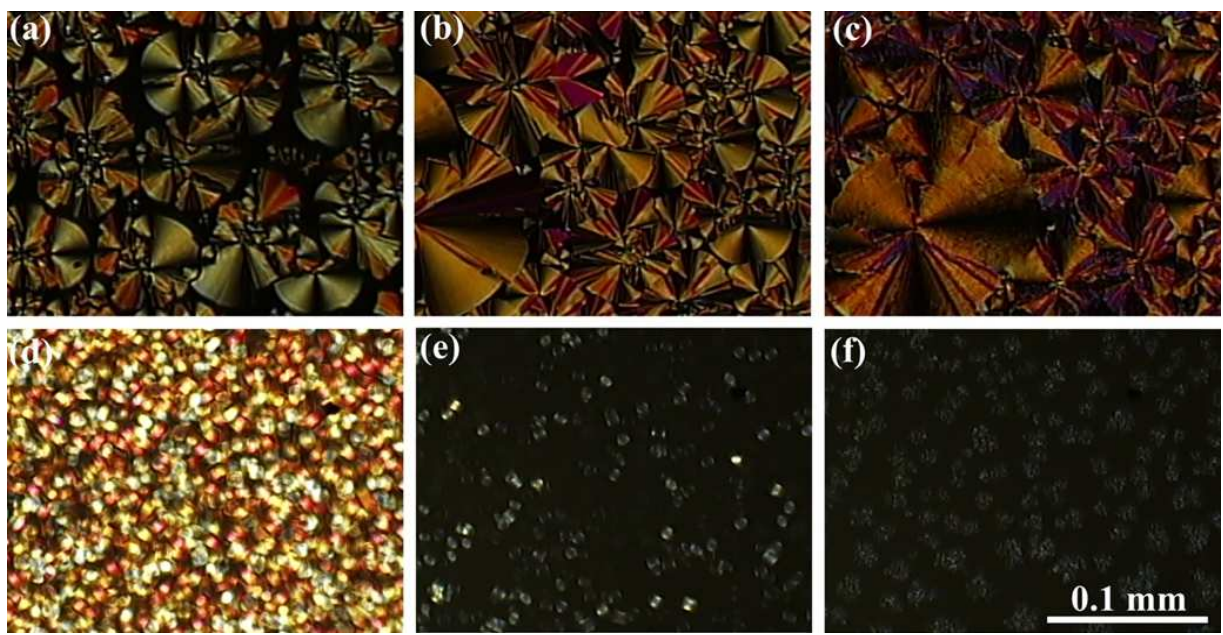

**Fig. 8** Polarizing optical microscopy images of 5 μm films of B5-Cl with polyimide inside substrates rubbed in horizontal direction at various conditions. In the top row the samples were illuminated only by microscope light. (a) 162 °C; (b) 103.5 °C; (c) 40 °C. Bottom row: samples at 130 °C after shining with UV light. (d) Before the second UV shining; (e) after the second UV shining; (f) the same as (e) but after keeping in dark for 15 hours.

## Conclusions

In continuation of our work on bent-core mesogens with pyridine as the central ring, we have synthesized a series of compounds with different substituents at the peripheral phenyl rings ($CH_3O$, Cl and $NO_2$). Comparing with the unsubstituted parent compound, the observed substituent effect on the mesomorphic behavior of the synthesized compounds demonstrates the delicate balance between the size and polarity of the substituent. UV light-induced changes in the texture and transition temperatures of the mesophase have also been observed. They might be attributed either to a partial decomposition or to a complex photoisomerization process.

Further work on the application of the investigated compounds is in progress.

## Experimental

### Preparations

The reagents were purchased from Sigma Aldrich and were used without further purification. All solvents were dried and distilled using general standard procedures. The chemical structures and the purities of the synthesized compounds were confirmed by elemental analysis, FT-IR, $^1$H and $^{13}$C NMR spectroscopy.

**Preparation of 4-dodecyloxy-3-substituted benzoic acids**

The synthesis of acids (**3a–3c**) was carried out according to the method described by Sivakumar *et al.*[29], while their mesomorphic properties have been presented elsewhere[35–37].

**3a**. FTIR (KBr) $v$ (cm$^{-1}$) = 3447, 2958, 2921, 2851, 2643, 2599, 2550, 1685, 1560, 1588, 1519, 1471, 1458, 1425, 1350, 1301, 1274, 1230, 1186, 1137, 1113, 1030, 942, 924, 877, 764, 642, 527; $^1$H NMR (200 MHz, CDCl$_3$) $\delta$: 10.54 (s, 1H, CO$_2$H), 7.75 (dd, $J$ = 8.5 and 1.7 Hz, 1H, Ar), 7.60 (d, $J$ = 1.8 Hz, 1H, Ar), 6.90 (d, $J$ = 8.6 Hz, 1H, Ar), 4.08 (t, $J$ = 6.8 Hz, 2H, CH$_2$O), 3.93 (s, 3H, CH$_3$O), 1.88 (m, 2H, CH$_2$), 1.60–1.13 (m, 18H, CH$_2$), 0.88 (t, $J$ = 6.6 Hz, 3H, CH$_3$); $^{13}$C NMR (50 MHz, CDCl$_3$): $\delta$ 172.2, 153.4, 148.9, 124.5, 121.3, 112.6, 111.3, 69.0, 56.0, 31.9, 29.6, 29.3, 28.9, 25.8, 22.6, 14.0. Elemental analysis, found: C, 71.32; H, 9.60. Calc. for C$_{20}$H$_{32}$O$_4$: C, 71.39; H, 9.59%.

**3b**. FTIR (KBr) $v$ (cm$^{-1}$) = 3416, 2953, 2923, 2850, 2642, 2590, 2534, 1687, 1596, 1568, 1504, 1465, 1430, 1386, 1323, 1301, 1278, 1150, 1127, 1062, 1020, 979, 938, 899, 821, 765, 716, 643, 571, 507; $^1$H NMR (200 MHz, CDCl$_3$) $\delta$: 9.49 (s, 1H, CO$_2$H), 8.12 (d, $J$ = 2.0 Hz, 1H, Ar), 7.98 (dd, $J$ = 8.6 and 2.2 Hz, 1H, Ar), 6.95 (d, $J$ = 8.8 Hz, 1H, Ar), 4.10 (t, $J$ = 6.5 Hz, 2H, CH$_2$O), 1.88 (m, 2H, CH$_2$), 1.58–1.13 (m, 18H, CH$_2$), 0.89 (t, $J$ = 6.6 Hz, 3H, CH$_3$); $^{13}$C NMR (50 MHz, CDCl$_3$): $\delta$ 171.1, 159.1, 132.3, 130.6, 122.9, 122.0, 112.0, 69.4, 31.9, 29.5, 29.3, 28.9, 25.9, 22.6, 14.1. Elemental analysis, found: C, 66.98; H, 8.54. Calc. for C$_{19}$H$_{29}$ClO$_3$: C, 66.95; H, 8.57%.

**3c**. FTIR (KBr) ν (cm$^{-1}$) = 3446, 2950, 2916, 2851, 2665, 2601, 2544, 1695, 1618, 1538, 1508, 1471, 1426, 1425, 1365, 1319, 1285, 1230, 1163, 1138, 1081, 996, 913, 829, 765, 716, 689, 543; $^1$H NMR (200 MHz, CDCl$_3$) δ: 10.51 (s, 1H, CO$_2$H), 8.39 (d, *J* = 2.2 Hz, 1H, Ar), 8.34 (dd, *J* = 8.8 and 2.2 Hz, 1H, Ar), 7.13 (d, *J* = 9.0 Hz, 1H, Ar), 4.18 (t, *J* = 6.4 Hz, 2H, CH$_2$O), 1.86 (m, 2H, CH$_2$), 1.60–1.02 (m, 18H, CH$_2$), 0.87 (t, *J* = 6.6 Hz, 3H, CH$_3$); $^{13}$C NMR (50 MHz, CDCl$_3$): δ 170.2, 156.4, 139.5, 135.7, 127.8, 121.1, 113.9, 70.2, 31.9, 29.6, 29.3, 28.7, 25.7, 22.6, 14.1. Elemental analysis, found: C, 64.95; H, 8.31; N, 3.45. Calc. for C$_{19}$H$_{29}$NO$_5$: C, 64.93; H, 8.32; N, 3.49%.

**Preparation of 2,6-bis[2-(4-(4-dodecyloxy-3-substituted benzoyloxy)phenyl)ethenyl]pyridines**

A mixture of 4-dodecyloxy-3-substituted benzoic acid (3 mmol), oxalyl chloride (15 mL) and few drops of DMF was refluxed for 6 hours. The excess oxalyl chloride was then removed and the obtained brownish solid was dried under vacuum for 2 hours. To a solution of the acid chloride in dry dichloromethane (20 mL), compound **B3** (0.47 g, 1.5 mmol) previously dispersed in dichloromethane (20 mL) was added, followed by the addition of triethylamine (0.41 mL, 3 mmol). The reaction mixture was stirred at the room temperature for 48 h. It was washed with 10% NaHCO$_3$ (2 × 10 mL) and several times with cold water and finally dried over anhydrous Na$_2$SO$_4$. The crude residue obtained after removal of solvent was chromatographed on silica gel using chloroform as eluent. Removal of solvent from the eluate afforded a white material which was further recrystallized from a hexane/toluene mixture.

**B5–MeO**. Yield: 60%; FTIR (KBr) ν (cm$^{-1}$) = 2955, 2921, 2851, 1731, 1599, 1561, 1514, 1469, 1455, 1417, 1385, 1344, 1274, 1214,1182, 1133, 1085, 1031, 973, 918, 875, 850, 804, 741, 524; $^1$H NMR (200 MHz, CDCl$_3$) δ: 7.82 (dd, *J* = 8.4 and 2.0 Hz, 2H, Ar), 7.76 (d, *J* = 17.0 Hz, 2H, CH=CH), 7.70 (d, *J* = 2.0 Hz, 2H, Ar), 7.69 (d, *J* = 8.8 Hz, 4H, Ar), 7.67 (t, *J* = 7.5 Hz, 1H, Py) - the signal at 7.67 is partly covered with the signal at 7.69, 7.29 (d, *J* = 7.6 Hz, 2H, Py), 7.26 (d, *J* = 8.0 Hz, 4H, Ar), 7.20 (d, *J* = 15.2 Hz, 2H, CH=CH), 6.97 (d, *J*= 8.6 Hz, 2H, Ar), 4.12 (t, *J* = 6.8 Hz, 4H, CH$_2$O), 3.97 (s, 6H, CH$_3$O), 1.92 (m, 4H, CH$_2$), 1.61–1.16 (m, 36H, CH$_2$), 0.90 (t, *J* = 6.4 Hz,6H, CH$_3$); $^{13}$C NMR (50 MHz, CDCl$_3$): δ 164.9, 155.3, 153.2, 151.0, 149.0, 136.9, 134.4, 131.9, 128.3, 128.1, 124.4, 122.1, 121.4, 120.6, 112.7, 111.4, 69.1, 56.1, 31.9, 29.6, 29.3, 28.9, 25.8, 22.6, 14.1. Elemental analysis, found: C, 76.87; H, 8.17; N, 1.48. Calc. for C$_{61}$H$_{77}$NO$_8$: C, 76.94; H, 8.15; N, 1.47%.

**B5–Cl**. The synthesis of **B5–Cl** was carried out according to the method described for **B5–MeO**. Yield: 70 %; FTIR (KBr) ν (cm$^{-1}$) = 2957, 2916, 2851, 1739, 1600, 1577, 1561, 1507, 1471, 1456, 1396, 1310, 1277, 1243, 1209, 1196, 1164, 1150, 1086, 1057, 1013, 969, 898, 842, 822, 753, 719, 630, 525; $^1$H NMR (200 MHz, CDCl$_3$) δ:8.25 (d, *J* = 2.2 Hz, 2H, Ar), 8.10 (dd, *J* = 8.6 and 2.2 Hz, 2H, Ar), 7.76 (d, *J* = 16.6 Hz, 2H, CH=CH), 7.70 (d, *J* = 8.8 Hz, 4H, Ar), 7.68 (t, *J* = 7.5 Hz, 1H, Py) ) – the signal at 7.68 is partly covered with the signal at 7.70, 7.30 (d, *J* = 7.6 Hz, 2H, Py), 7.26 (d, *J* = 8.8 Hz, 4H, Ar) – the signal at 7.26 is partly covered with the signal at 7.30, 7.21 (d, *J* = 16.0 Hz, 2H, CH=CH), 7.02 (d, *J* = 8.8 Hz, 2H, Ar), 4.15 (t, *J* = 6.0 Hz, 4H, CH$_2$O),1.92 (m, 4H, CH$_2$), 1.62–1.15 (m, 36H, CH$_2$), 0.92 (t, *J* = 6.5 Hz,6H, CH$_3$); $^{13}$C NMR (50 MHz, CDCl$_3$): δ 163.9, 158.9, 155.3, 150.8, 137.0, 134.6, 132.2, 131.9, 130.5, 128.5, 128.2, 123.0, 122.1, 122.0, 120.6, 112.2, 69.4, 31.9, 29.6, 29.3, 28.8, 25.8, 22.6, 14.1. Elemental analysis, found: C, 73.75; H, 7.43; N, 1.47. Calc. for C$_{59}$H$_{71}$ClNO$_6$: C, 73.73; H, 7.45; N, 1.46%.

**B5–NO$_2$**. The synthesis of **B5–NO$_2$** was carried out according to the method described for **B5–MeO**. Yield: 45%; FTIR (KBr) ν (cm$^{-1}$) = 2957, 2922, 2851, 1740, 1624, 1561, 1538, 1507, 1470, 1456, 1384, 1347, 1292, 1274, 1246, 1211, 1153, 968, 923, 910, 830, 813, 751, 698, 649, 629, 523; $^1$H NMR (200 MHz, CDCl$_3$) δ:8.66 (d, *J* = 2.2 Hz, 2H, Ar), 8.33 (dd, *J* = 9.0 and 2.2 Hz, 2H, Ar), 7.74 (d, *J* = 15.6 Hz, 2H, CH=CH), 7.68 (d, *J* = 8.2 Hz, 4H, Ar), 7.66 (t, *J* = 7.2 Hz, 1H, Py) – the signal at 7.66 is partly covered with the signal at 7.68,7.28 (d, *J* = 7.2 Hz, 2H, Py), 7.24 (d, *J* = 7.4 Hz, 4H, Ar), 7.19 (d, *J* = 15.4 Hz, 2H, CH=CH), 7.17 (d, *J* = 9.2 Hz, 2H, Ar), 4.20 (t, *J* = 6.4 Hz, 4H, CH$_2$O),1.88 (m, 4H, CH$_2$), 1.61–1.15 (m, 36H, CH$_2$), 0.88 (t, *J* = 6.4 Hz,6H, CH$_3$); $^{13}$C NMR (50 MHz, CDCl$_3$): δ 163.9, 158.9, 155.3, 150.8, 137.0, 134.6, 132.2, 131.9, 130.5, 128.5, 128.2, 123.0, 122.1, 122.0, 120.6, 114.0, 70.2, 31.9, 29.6, 29.3, 28.4, 25.7, 22.6, 14.1. Elemental analysis, found: C, 72.12; H, 7.28; N, 4.31. Calc. for C$_{59}$H$_{71}$N$_3$O$_{10}$: C, 72.15; H, 7.29; N, 4.28%.

**Crystal structure determination**

Single crystals of suitable size were grown by a slow evaporation of acetonitrile solution in a refrigerator. Diffraction data are collected at 293 K on an Oxford Gemini S diffractometer equipped with CCD detector using monochromatized Mo *K*α radiation (λ = 0.71073 Å). Intensities were corrected for absorption by means of the multi-scan method. The structure was solved by direct methods (SIR92)[38] and refined on *F*$^2$ by full-matrix least-squares using the programs SHELXL-2014/7[39] and WinGX[40]. All non-hydrogen atoms were refined anisotropically.

**Table 4** Crystal data and structure refinement for **B3**

| Formula | C$_{21}$H$_{17}$NO$_2$ |
|---|---|
| Formula mass (g mol$^{-1}$) | 315.36 |
| Crystal system | Monoclinic |
| Space group | P2$_1$/n |
| *a* (Å) | 10.9312(6) |
| *b* (Å) | 9.1547(9) |
| *c* (Å) | 16.4772(9) |
| β(°) | 93.129(5) |
| *V* (Å$^3$) | 1646.4(2) |
| *Z* | 4 |
| ρ$_c$ (g cm$^{-3}$) | 1.272 |
| μ (mm$^{-1}$) | 0.082 |
| *F*(000) | 664 |
| Crystal size (mm) | 0.3 × 0.1 × 0.03 |
| θ range (°) | 2.904 – 25.349 |
| Index ranges | −13≤*h*≤ 11 |

| | −11 ≤ k ≤ 8 |
| --- | --- |
| | −19 ≤ l ≤ 19 |
| Reflections collected | 7529 |
| Independent reflections | 3006 ($R_{int}$ = 0.0837) |
| Data / restraints / parameters | 3006 / 0 / 217 |
| Goodness-of-fit on $F^2$, S | 1.098 |
| R indices [I > 2σ(I)] | $R_1$ = 0.0975 |
| | $wR_2$ = 0.1125 |
| R indices (all data) | $R_1$ = 0.1946 |
| | $wR_2$ = 0.1352 |
| $\Delta\rho_{max}$, $\Delta\rho_{min}$ (e Å$^{-3}$) | 0.180, −0.172 |

Positions of H atoms were calculated on geometric criteria and refined using the riding model with $U_{iso}$(H) = 1.2$U_{eq}$(C) and $U_{iso}$(H) = 1.5$U_{eq}$(O). Selected crystal data and refinement results are listed in Table 4. Crystallographic data are deposited at the Cambridge Crystallographic Data Centre with CCDC reference number 1448109.

**Physicochemical investigations**

To confirm the chemical structure of the synthesized compounds, the following analytical methods were applied. Elemental analysis was realized using an Elemental Vario EL III microanalyzer, their results were found to be in good agreement (±0.3%) with the calculated values. FT-IR spectra were recorded on a Bomem MB 100 spectrophotometer ($v_{max}$ in cm$^{-1}$) on KBr pellets. $^1$H and $^{13}$C NMR spectra were recorded on a Varian Gemini 200 spectrometer at 200 MHz for $^1$H NMR and 50 MHz for $^{13}$C NMR spectra. NMR spectra were recorded in CDCl$_3$, or DMSO-$d_6$, using TMS as the internal standard (chemical shift δ in ppm).

For POM observations presented in Fig. 6, Zeiss Axio Imager.A1 polarizing microscope has been used, equipped with Instec HS-1 hot-stage (having a thermal stability of 2mK) and with an Optronics CCD camera. The temperatures and enthalpies of the phase transitions were determined by differential scanning calorimetry on a Mettler FP 800HT with FP85TA using cooling and heating runs at a rate of 5 °C min$^{-1}$. The samples of 3–5 mg were sealed in aluminium pans.

Small angle x-ray scattering measurements were obtained in Advance Light Source of Lawrence Berkeley National Laboratory on beamline 7.3.3 (10 keV incident beam energy (1.24 Å wavelength), utilizing a Pilatus 2M detector)[41]. The materials were filled into 1mm diameter quartz X-ray capillary tubes, which were then mounted into a custom-built aluminium cassette that allowed X-ray detection with ±13.5° angular range. The cassette fits into a standard hot stage (Instec model HCS402) that allowed temperature control with ±0.1 °C precision. The stage also included two cylindrical neodymium iron boron magnets that supplied a magnetic induction of B=1.5T perpendicular to the incident X-ray beam. The analysis were proceed in Igor Pro software with Nika package[42].

## Acknowledgements

This work was supported by the Ministry of Education, Science and Technological Development of the Republic of Serbia (Projects no. 172013 and III45007). T.T.-K. and K.F.-Cs. acknowledge the financial support of EU FP7 M-era.Net -- MACOSYS project (Hungarian Research Fund OTKA NN 110672).AJ and MS acknowledge useful help and discussions with James T. Gleeson and Samuel Sprunt from the Department of Physics of Kent State University, and Chenhui Zhu at the Advanced Light Source of Lawrence Berkeley National Laboratory, and financial support from NSF DMR-1307674. The Advanced Light Source is supported by the Director, Office of Science, Office of Basic Energy Sciences, of the U.S. Department of Energy under Contract No. DE-AC02-05CH11231.